\documentclass[aps,superscriptaddress,showpacs,twocolumn]{revtex4-1}
\usepackage{amsmath,amssymb,graphics,graphicx,dcolumn,bm,enumerate}
\usepackage{multirow}

\newcommand{\ictp}
{\affiliation{CMSP Section,
The Abdus Salam International Centre for Theoretical Physics,
Strada Costiera 11, Trieste I-34014, Italy.}}
\newcommand{\sissa}
{\affiliation{International School for Advanced Studies (SISSA) and INFN, Via Bonomea 265, I-34136 Trieste, Italy.}}
\newcommand{\sinp}
{\affiliation{TCMP Division,
Saha Institute of Nuclear Physics, 1/AF Bidhannagar, Kolkata 700 064, India.}}
\newcommand{\camcs}
{\affiliation{Centre for Applied Mathematics and Computational Science,\\
Saha Institute of Nuclear Physics, 1/AF Bidhannagar, Kolkata 700 064, India.}}
\newcommand{\isi}
{\affiliation{Economic Research Unit, Indian Statistical Institute, 203 B. T. Road, Kolkata 700 018, India}}
\newcommand{\cpt}
{\affiliation{Centre de Physique Th\'{e}orique (CNRS UMR 6207),\\
 Universit\'{e} de la M\'{e}diterran\'{e}e Aix Marseille II,
Luminy, 13288 Marseille cedex 9, France.}}
\newcommand{\roma}
{\affiliation{Dipartimento di Fisica, Sapienza Universit\'a di Roma , P.le A. Moro 2, 00185, Rome, Italy}}

\begin{document}
\title{Phase transitions in crowd dynamics of resource allocation}

\author{Asim Ghosh}
\email[Email: ]{asim.ghosh@saha.ac.in}
\sinp
\author{Daniele De Martino}
\email[Email: ]{daniele.demartino@roma1.infn.it}
\sissa \roma
\author{Arnab Chatterjee}
\email[Email: ]{arnab.chatterjee@cpt.univ-mrs.fr}
\ictp \cpt
\author{Matteo Marsili}
\email[Email: ]{marsili@ictp.it}
\ictp
\author{Bikas K. Chakrabarti}
\email[Email: ]{bikask.chakrabarti@saha.ac.in}
\sinp \camcs \isi

\begin{abstract}
We define and study a class of resources allocation processes where $gN$ agents, by
repeatedly visiting $N$ resources, try to converge to optimal configuration where each resource is 
occupied by at most one agent. The process exhibits a phase transition, as the density $g$ of agents grows, 
from an absorbing to an active phase. In the latter, even if the number of resources is in principle enough for all agents ($g<1$), 
the system never settles to a frozen configuration. 
We recast these processes in terms of zero-range interacting particles, 
studying analytically the mean field dynamics 
and investigating numerically the phase transition in finite dimensions. 
We find a good agreement with the critical exponents 
of the stochastic fixed-energy sandpile. 
The lack of coordination in the active phase also leads to a non-trivial faster-is-slower effect.     
\end{abstract}


\maketitle

\section{Introduction}
\noindent A general question that naturally arises in the study of social systems 
is how collective behaviors can emerge from the interactions among individuals. 
In spite of the inherent complexity of these phenomena, simplified mathematical models 
that assume simple automatic responses of individuals to stimuli 
can reproduce non-trivial effects in the observed behavior~\cite{ball}.
Statistical mechanics had a certain success  
in the use of coarse-grained models of physical systems 
to connect the microscopic dynamics with the macroscopic behavior, 
and its techniques and concepts are starting to be fruitfully applied 
to understand social dynamics~\cite{Liggett:1985,Castellano:2009,EconoSocio2006,vegaredon}.
In particular, the phenomena observed in the crowd dynamics, 
from pedestrians flows~\cite{Helbing:2009} to vehicular traffic~\cite{Helbing:2001}, 
have been recently subject to quantitative measures.
Observations range from bottleneck oscillations, lanes and stripes formation, intermittent flows,
waves, turbulence~\cite{Helbing:2007}, faster-is-slower~\cite{Helbing:2000a} and freezing-by-heating effects~\cite{Helbing:2000b}.
Interestingly, this extremely broad class of collective phenomena is explained at semi-quantitative level  
with the use of models of interacting particles or granular fluids.
This shows that even if interactions among individuals in these settings 
are rather simplified and mechanical, they still lead to a certain level of collective coordination. 

In this paper we show how a crowd dynamics in a resources allocation context can give rise to a phase transitions between an active  
and an absorbing phase. The class of models we discuss is inspired by  the 
Kolkata Paise Restaurant (KPR) problem~\cite{Chakrabarti:2009,mathematica,proc,Ghosh:2010}, which we generalize and recast in 
the broader context of zero-range interacting particle models \cite{Evans:2005}. 

The KPR is a repeated game played by a large number ($gN$, $g$ being a real number) of agents with equally shared past information. 
Every evening they try to get the best service in one of the $N$ restaurants  of commonly agreed ranks, 
each providing food for one person only. 
It serves as a paradigm for a problem of resource utilization, where agents 
learn from their actions to maximize the effective utilization of available resources. 
At unit density ($g=1$), a simple random choice algorithm~\cite{Chakrabarti:2009} leads 
(in unit convergence time) to a occupation of a fraction of $f=1-e^{-1} \simeq 0.63$ only of the resources, 
which of course falls much short of a fully efficient usage of them. As seen in various earlier  
studies~\cite{mathematica,proc,Ghosh:2010}, apparently smarter strategies fail to yield better results. However, a stochastic 
strategy~\cite{Ghosh:2010} that maintains a na\"{i}ve tendency 
to stick to an agent's past choice, with probability decreasing with the 
crowd size, leads to an efficient utilization fraction $f$ of about $0.796$~\cite{Ghosh:2010}  
and it converges to the above within a time (iteration) independent of $N$. 
An extension of this strategy to the Minority game~\cite{MGBook:2005} problem
(with knowledge of both the sign and size of the population difference among two alternative choices), 
gives the optimal solution (with fluctuation bounded by $\log N$ and convergence 
time bounded by $\log (\log N)$~\cite{deepak}).
Our focus will be on the nature of the collective behavior rather than on the efficiency of individual behavior.
For the latter, we refer to applications of game theory and adaptive learning, as e.g. 
in~\cite{Arthur:1994,Kandori,Nowak:1993,Orlean:1995,Hanaki:2011}. 
We note, in passing, that the relation between the degree of individual rationality and the efficiency of collective allocation can be a non-trivial one, as shown e.g. in Refs.~\cite{probstrat,Satinover:2007}.

Here we will study the general problem with $N$ restaurants 
and $g N$ agents, where $g$, the density, is a fixed external parameter of the dynamics.
Recasting the problem in terms of zero-range interacting particles, 
we report a phase transition from a frozen phase with satisfied agents to an active phase 
with unsatisfied ones at a critical density $g_c$ and perform extensive numerical simulations 
as well as some analytical calculations to understand its features, finding a good 
agreement with the exponents of stochastic fixed-energy sandpiles \cite{Vespignani}.
The study of the relaxation properties of the frozen phase reveals 
an interesting faster-is-slower effect:
This consists in the general observation that, from vehicular traffic to logistic, 
a higher level of coordination can arise from 
strategies which involve a slower dynamics (e.g. waiting), that however speed up performances.
   
The paper is organized as follows: In Sec.~\ref{sec:model}, we introduce the class of models
under study. Next, in Sec.~\ref{sec:simul} we present the results of numerical simulations 
and in Sec.~\ref{sec:ana} we provide the analytic treatment for the mean field case.
We conclude with a summary and some discussions in Sec.~\ref{sec:sum}.

\begin{figure*}
\includegraphics[width=5.9cm]{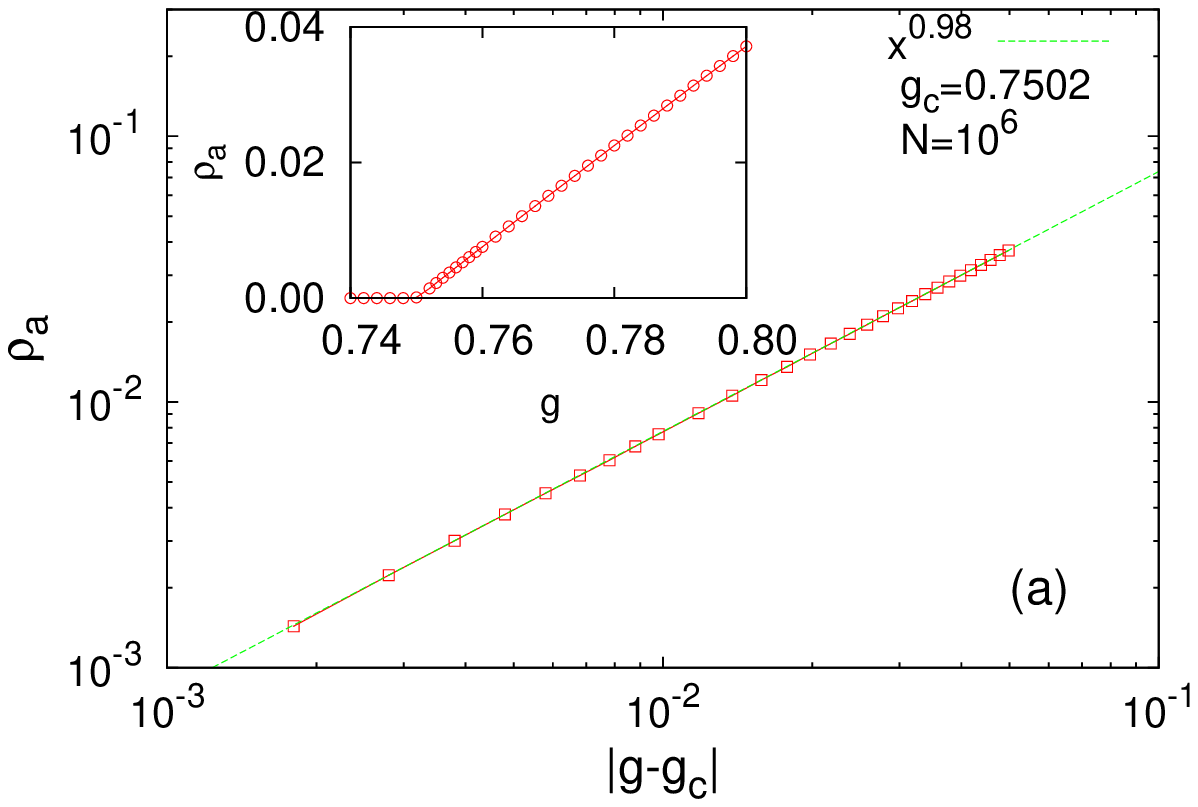}
\includegraphics[width=5.9cm]{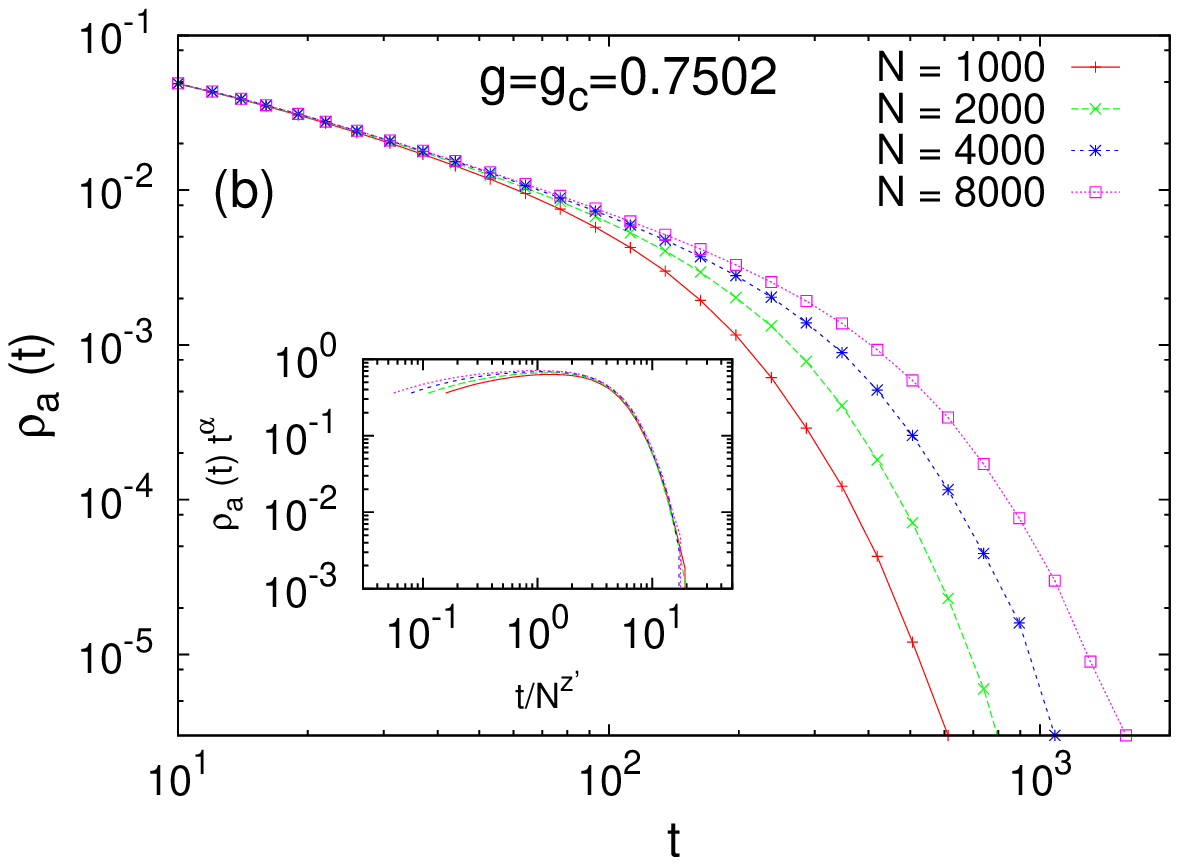}
\includegraphics[width=5.9cm]{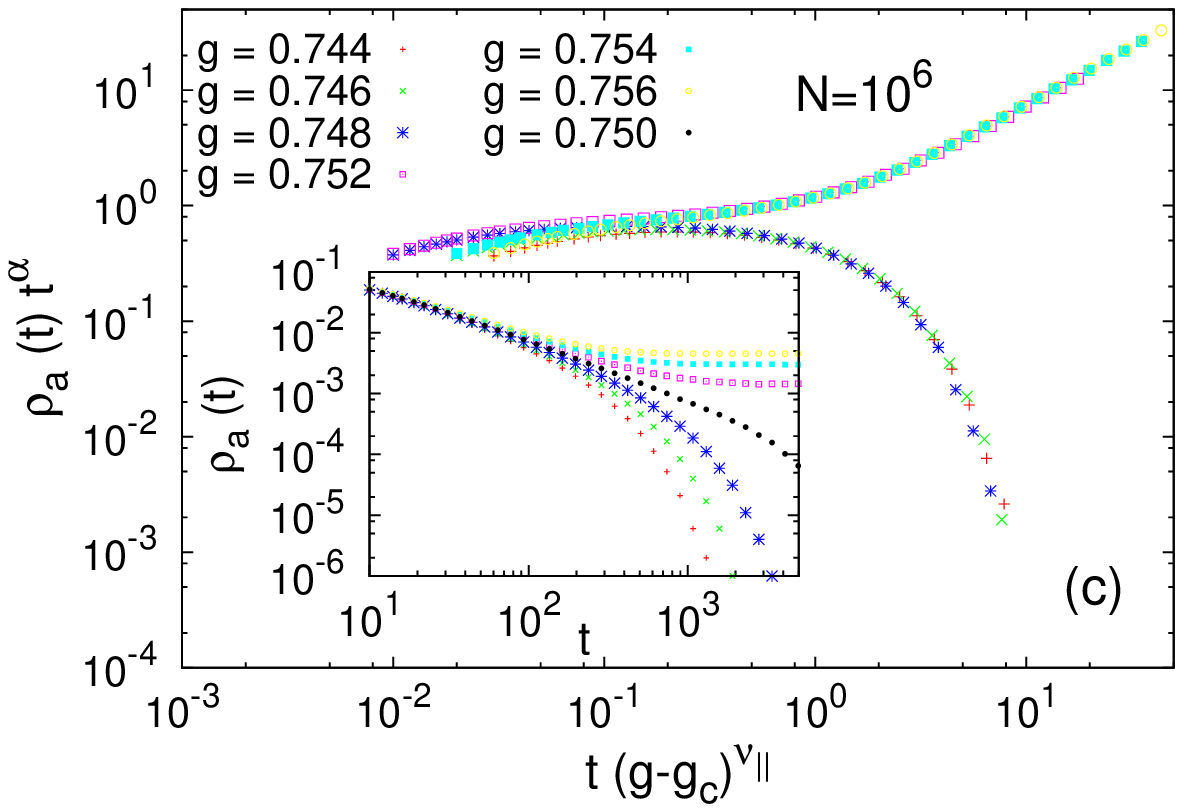}
\caption{Simulation results for mean field case, $g_c = 0.7502 \pm 0.0002$.
(a) Variation of steady state density $\rho_a$ of active sites versus $g-g_c$, fitting to $\beta=0.98 \pm 0.02$.
The inset shows the variation of $\rho_a$ with density $g$.
(b) relaxation to absorbing state near critical point for different system sizes,
the inset showing the scaling collapse giving estimates of critical
exponents $\alpha=1.00 \pm 0.01$ and $z^{\prime}=0.50\pm 0.01$. 
(c) Scaling collapse of $\rho_a(t)$. 
The inset shows the variation of $\rho_a(t)$ versus time $t$ for different densities $g$.
The estimated critical exponent is $\nu_\parallel = 1.00 \pm 0.01$.
The system sizes $N$ are mentioned.}
\label{fig:mfn}
\end{figure*}
\begin{figure*}
\includegraphics[width=5.9cm]{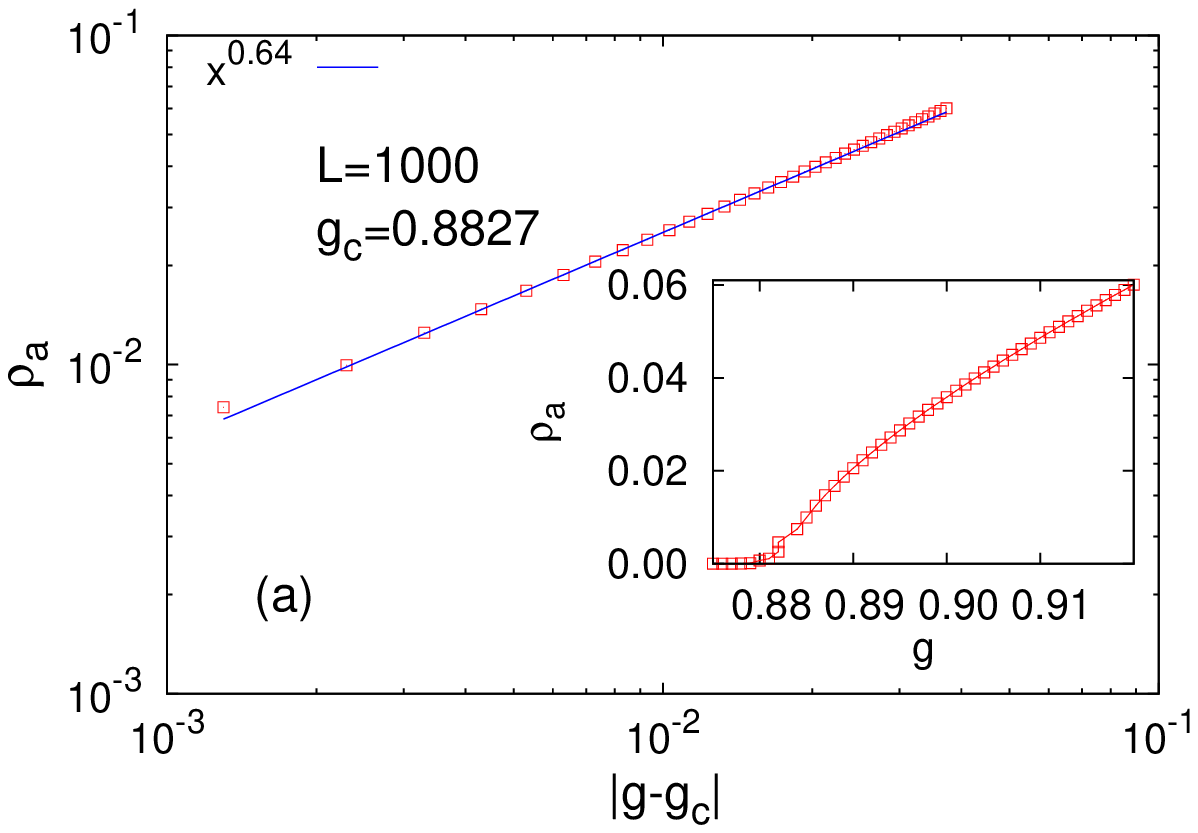}
\includegraphics[width=5.9cm]{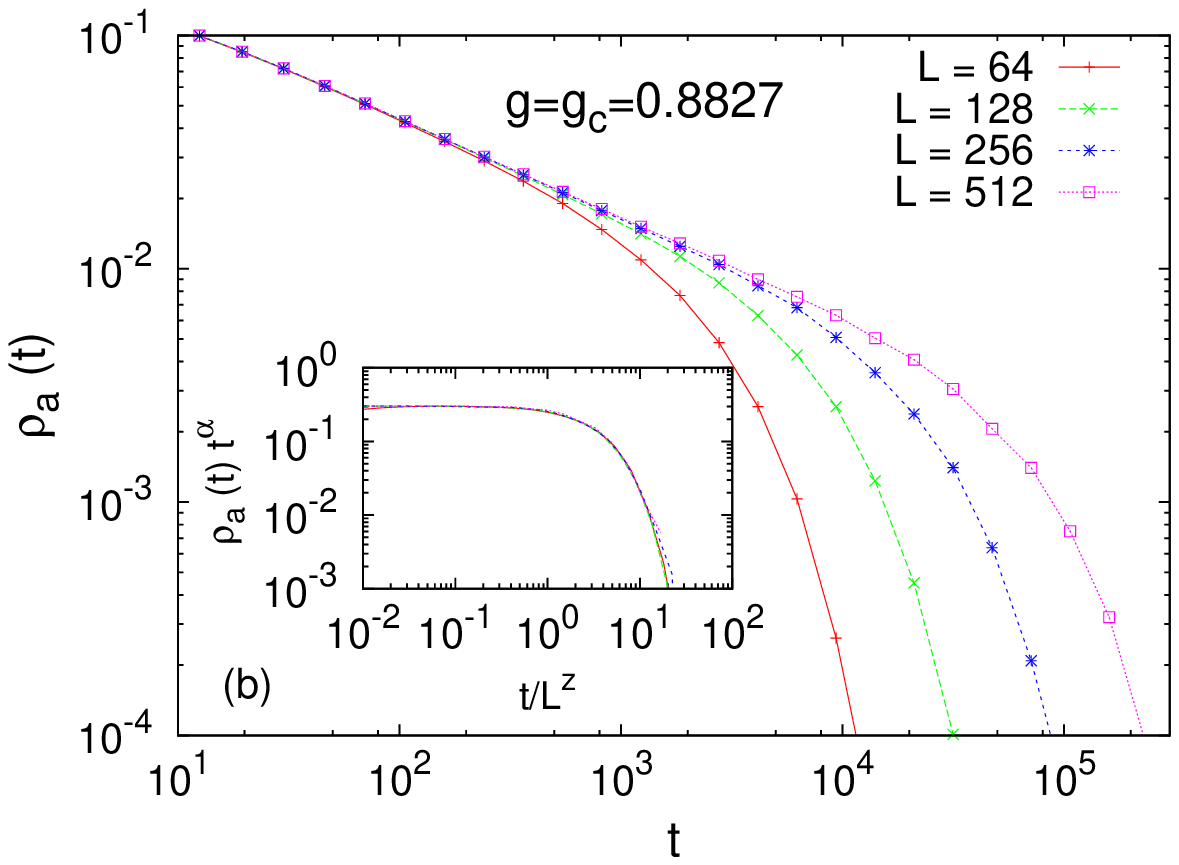}
\includegraphics[width=5.9cm]{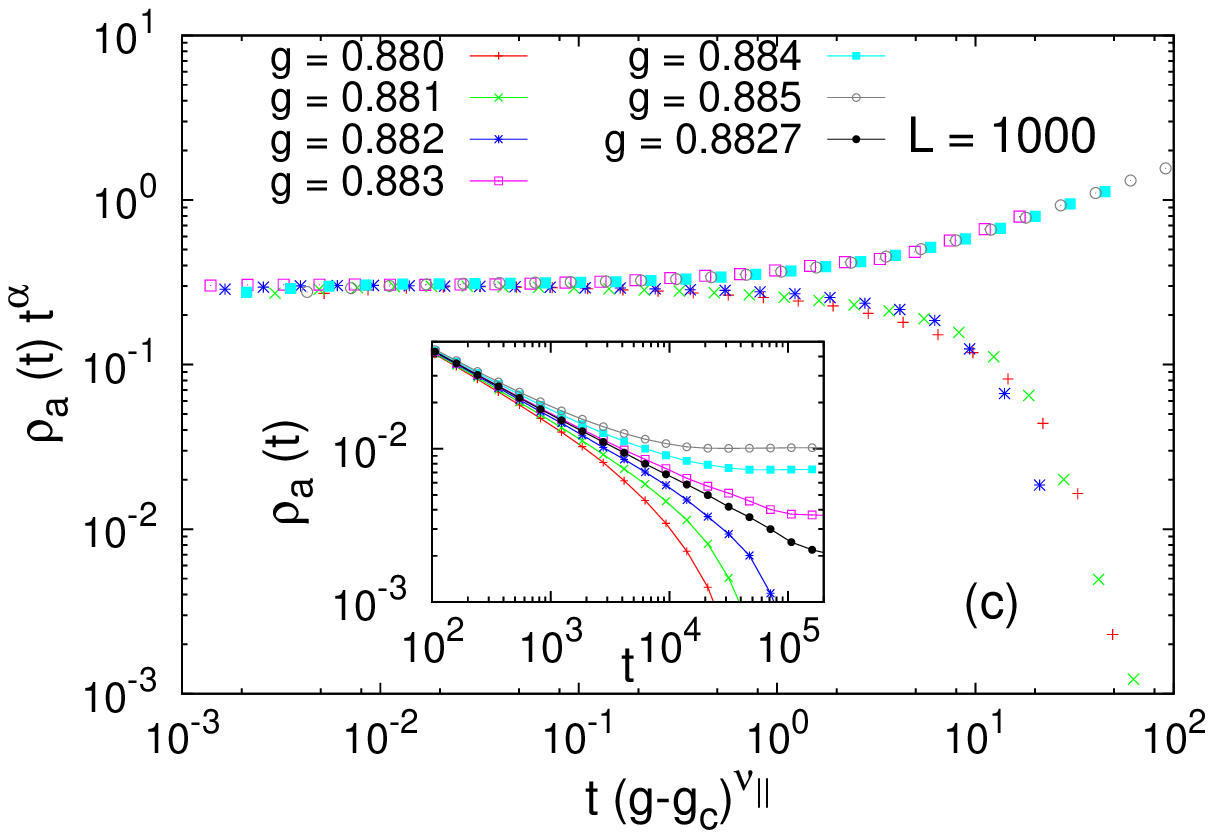}
\caption{Simulation results for $2$-d case, $g_c = 0.8827 \pm 0.0002$.
(a) Variation of steady state density $\rho_a$ of active sites versus $g-g_c$, fitting to $\beta=0.68 \pm 0.01$.
The inset shows the variation of $\rho_a$ with density $g$.
(b) relaxation to absorbing state near critical point for different system sizes,
the inset showing the scaling collapse giving estimates of critical
exponents $\alpha=0.42 \pm 0.01$ and $z=1.65\pm 0.02$. 
(c) Scaling collapse of $\rho_a(t)$. 
The inset shows the variation of $\rho_a(t)$ versus time $t$ for different densities $g$.
The estimated critical exponent is $\nu_\parallel = 1.24 \pm 0.01$.
The simulations are done for square lattices of linear size $L$ ($N=L^2$).
}
\label{fig:2d1n}
\end{figure*}

\section{The models}
\label{sec:model}
\noindent Inspired by the KPR problem~\cite{Chakrabarti:2009,Ghosh:2010} with $N$ individuals 
competing for $N$ restaurants (each serving food to one person each), 
we propose a more generic and fundamental stochastic occupation problem with exclusion.
 
The rank ordering among the restaurants, as in the original KPR 
problem~\cite{Chakrabarti:2009,Ghosh:2010} is dropped here and we consider in general $g N$ agents, 
where $g$, the density, is an external fixed parameter of the dynamics.
We will refer from now on to individuals as particles 
and restaurants as the sites or nodes of an underlying graph. 
In these terms the original problem is defined in a fully connected geometry. 

A particle moves from the site $i$ it occupies to a randomly chosen neighbor $j$ with a rate that depends 
only on the number of particles that is present on it $v(n_i)$. 
This is by definition a {\em zero range process} (see \cite{Evans:2005} for a review) which allows us to 
conclude that the stationary probability distribution of the number of particles per site 
factorizes in terms of single site functions.  
Given the nature of the problem, we will consider models with 
$v(1)=0$, i.e. costumers are happy while alone, and $v(n+1)\geq v(n)$, 
i.e. repelling particles. Given this expression for the rates, at low densities ($g<1$) 
there are dynamically frozen configurations with sites filled by single particles.  
On the other hand, for  high densities ($g>1$) a finite fraction of sites -- that we shall call  `active' -- will have multiple occupancy.
We will show that there is a  transition between these two phases that occurs at a certain density $g_c\le 1$. 
Notice that, in principle, the process is ergodic so every configuration of is accessible. Therefore, for $g\le 1$, the process will sooner or later visit an absorbing configuration where $n_i\le 1$ for all sites $i$. Still, when $N$ is large enough and $g>g_c$, frozen configurations will be visited extremely rarely, over time-scales which are much longer than those which are accessible to numerical simulations.
We choose as the order parameter the steady state density 
of active sites $\rho_a$ (density of sites having $n>1$). So the absorbing phase corresponds to $\rho_a=0$ whereas, above 
some density  $g_c$ the steady state shows a persistent dynamics with a non-zero value of the order parameter ($\rho_a > 0$).

We will analyze in particular two models:
\begin{itemize}
\item[(A)] $v(n) = 1-\frac{1}{n}$ 
\item[(B)] $v(n) = (1-p) \theta(n-1)$.
\end{itemize}

We will refer from now on to a parallel dynamics in which a simultaneous update of the nodes is done at each time step,
i.e. agents' actions are simultaneous akin to a repeated game setting. 

It  will be pointed out later that a sequential update 
in which at each step a randomly chosen particle jumps with some probability leads here to the trivial result that $g_c=1$. This corresponds to the limit $p\to 1$ of model B.

The model A implements the stochastic crowd avoiding strategy of the original KPR problem\cite{Ghosh:2010}. 
Here, if the site $k$ has $n_k \ge 1$ particles, each of them stay back with probability
 $1/n_k$ in the next time step, otherwise it jumps to any of the neighboring sites.  

In the model B an external parameter $p$ is introduced that represents 
the ``patience'' of costumers to overcrowded conditions.  
Here, if the site $k$ has $n_k \ge 1$ particles, each of them stays with probability
$p$ in the next time step, otherwise it jumps to any of the neighboring sites.  

The model B is in practice a kind of fixed energy sandpile  \cite{Vespignani}, 
but the study of its dynamics as a function of the parameter 
$p$ will reveal an interesting faster-is slower effect 
related to the relaxation time of the frozen phase.

We finally point out that a waiting choice can be {\em rational} from the point of view of game theory:
the agents in overcrowded sites could wait simply because they expect that others are leaving them alone.

In the next section we report a detailed analysis of the phase transition of 
both models with the use of numerical simulations in a fully connected geometry, 
a 1d chain and a 2d square lattice. Then, we perform analytical calculations for the mean field case.

\section{Results from numerical simulations}
\label{sec:simul}
\noindent We measure the times required to reach the steady state below and above $g_c$. 
Below $g_c$, the order parameter $\rho_a$ reaches a value $\rho_a =0$ in the steady state.  
Above $g_c$,  order parameter $\rho_a$ evolves to a stationary state 
and fluctuates around a mean value $\rho_a^0$  $(> 0)$. The system has persistent dynamics 
in this phase. The evolution of the order parameter is exponential away from $g_c$, and can be
expressed as
\begin{equation}
\label{tau+def1}
 \rho_a(t)  = \rho_a^0 \left[ 1 - e^{-t/\tau}\right]
\end{equation}
for $g>g_c$, and
 \begin{equation}
\label{tau+def2}
     \rho_a(t)  \propto e^{-t/\tau}
\end{equation}
for $g<g_c$, where $\tau$ is the relaxation time. We will denote the asymptotic value
of the order parameter as $\rho_a$ hereafter.
Near the critical point ($g-g_c \to 0_+$), 
we find  $\rho_a \sim (g - g_c)^\beta$ where $\beta$ is the order parameter exponent,
and $\tau \sim (g -g_c)^{-\nu_\parallel}$. Generally  $\rho_a(t)$  obeys a scaling form
\begin{equation}
\label{scaling-relation}
 \rho_a(t) \sim t^{-\alpha} \mathcal{F}\left( \frac{t}{\tau} \right);~~
\tau\sim(g-g_c)^{-\nu_\parallel}\sim L^z,
\end{equation}
where $\alpha$ and $z$ are dynamic exponents and $L$ denotes size of the system. 
We then get $\beta=\nu_\parallel \alpha$ by comparing 
Eq.~(\ref{tau+def1}), Eq.~(\ref{tau+def2}) and Eq.~(\ref{scaling-relation}) when $t/\tau$ 
is a constant for $t\rightarrow \infty$.  We study numerically the 
time variation of $\rho_a(t)$ and measure the exponents by  fitting to the above scaling relation. 

\subsection{Model A}

\subsubsection{Mean Field case}
\noindent For the mean field case, we have studied systems of $N=10^6$ sites, averaging over $10^3$ initial conditions. 
We get $g_c =0.7502 \pm 0002$. 
The scaling fits of $\rho_a(t)$ for different $g$ values (see Fig.~\ref{fig:mfn}) give 
$\beta = 0.98 \pm0.02 $, $z^{\prime} = 0.50 \pm 0.01$ (if we assume $N=L^4$ and using Eqn. (\ref{scaling-relation}), we get a relation $z=4z^{\prime}$  and therefore $z=2.0\pm0.04$), $\nu_\parallel = 1.00 \pm 0.01$, 
$\alpha = 1.00 \pm 0.01$.  
It may be noted that these independently estimated exponent values satisfy 
the scaling relation $\beta=\nu_\parallel \alpha$ well. 


\subsubsection{Lattice cases}
\noindent We studied the same dynamics in $1$-d and $2$-d. For a linear chain in $1$-d, 
we took $N=L=10^4$ and averaged over $10^3$ initial conditions. 
For $2$-d we consider square lattice ($N=L^2$) with $L=1000$ and averaging over 
$10^3$ initial conditions. Periodic boundary condition were employed in both cases.
\begin{enumerate}[(a)]
\item The model is defined in $1$-d
as follows: The particles are allowed to hop only to their nearest neighbor sites, 
and each particle can choose either left or right neighbor randomly.  
We find  $g_c = 1$ and hence the phase transition is not very interesting.
\item In the $2$-d version of the model, we consider square lattices 
and the particles are allowed to choose one of the $4$ nearest neighbors randomly.  
For $N=1000 \times 1000$, we get  $g_c = 0.88 \pm 0.01$, $\beta = 0.68 \pm 0.01$, 
$z = 1.65 \pm 0.02$, $\nu_\parallel = 1.24 \pm 0.01$ and  
$\alpha = 0.42 \pm 0.01$ (Fig.~\ref{fig:2d1n}). 
It may be noted that these independently estimated exponent values do not 
fit very well with the scaling relation $\beta=\nu_\parallel \alpha$. 
However, this type of scaling violation is also obseved in many active-absorbing transition 
cases~\cite{Rossi:2000}.

\end{enumerate}
\begin{figure}
\centering
\includegraphics[width=8.7cm]{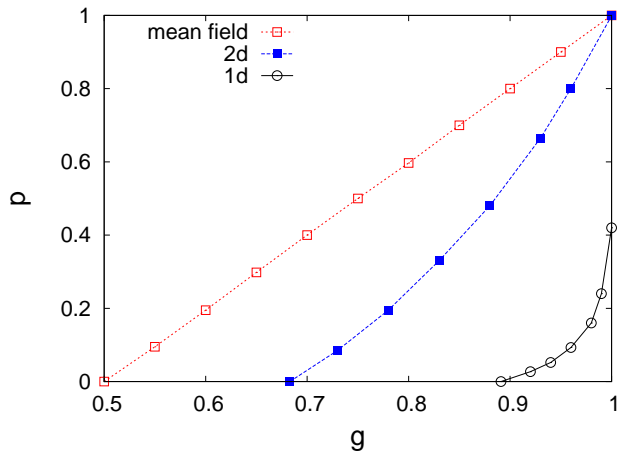}
\caption{Phase diagram for the generalized model in the  $(g,p)$ plane, 
showing the phase boundaries separating the active and absorbing phases
 in 1-d, 2-d and mean field cases. 
The active phases are on the right of the phase boundaries
while the absorbing phases are on the left in the respective cases.
The system sizes are $N=10^5$ for mean field, $1000 \times 1000$ for $2$-d, and $10^4$ for $1$-d.}
\label{fig:pg}
\end{figure}

\begin{figure*}
\centering
\includegraphics[width=5.9cm]{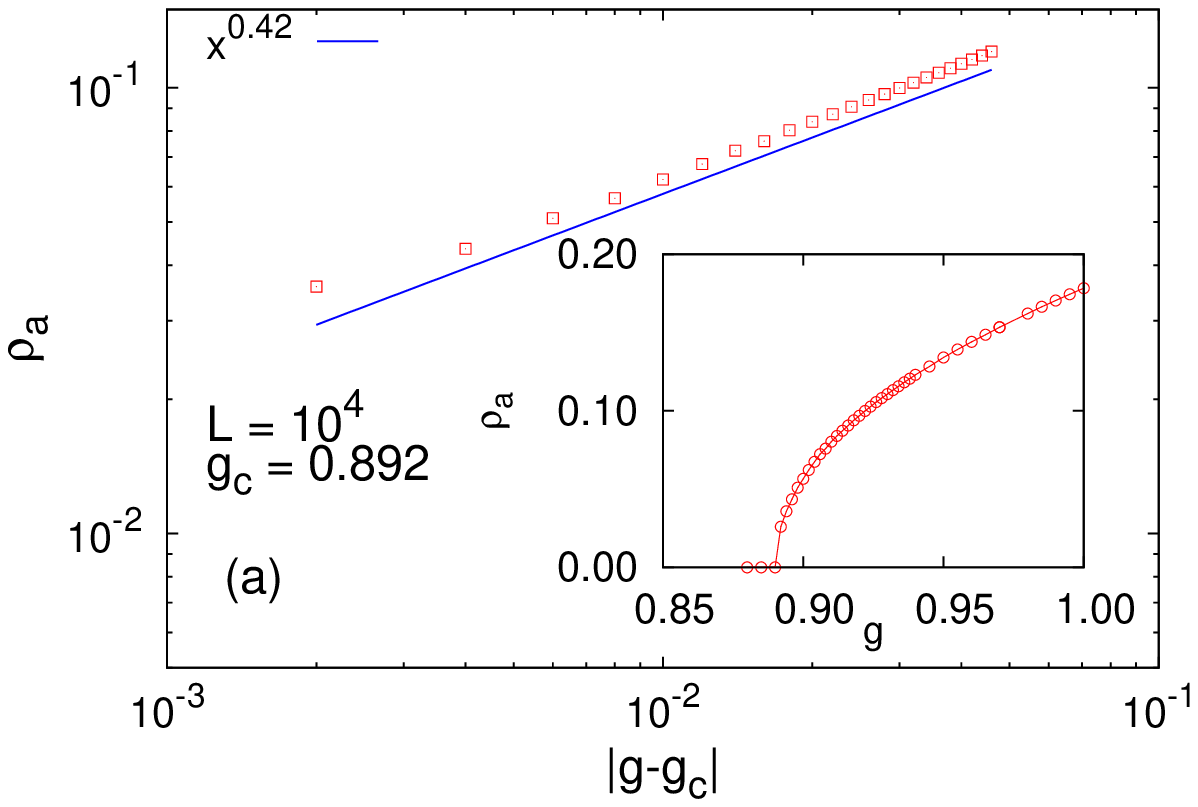}
\includegraphics[width=5.9cm]{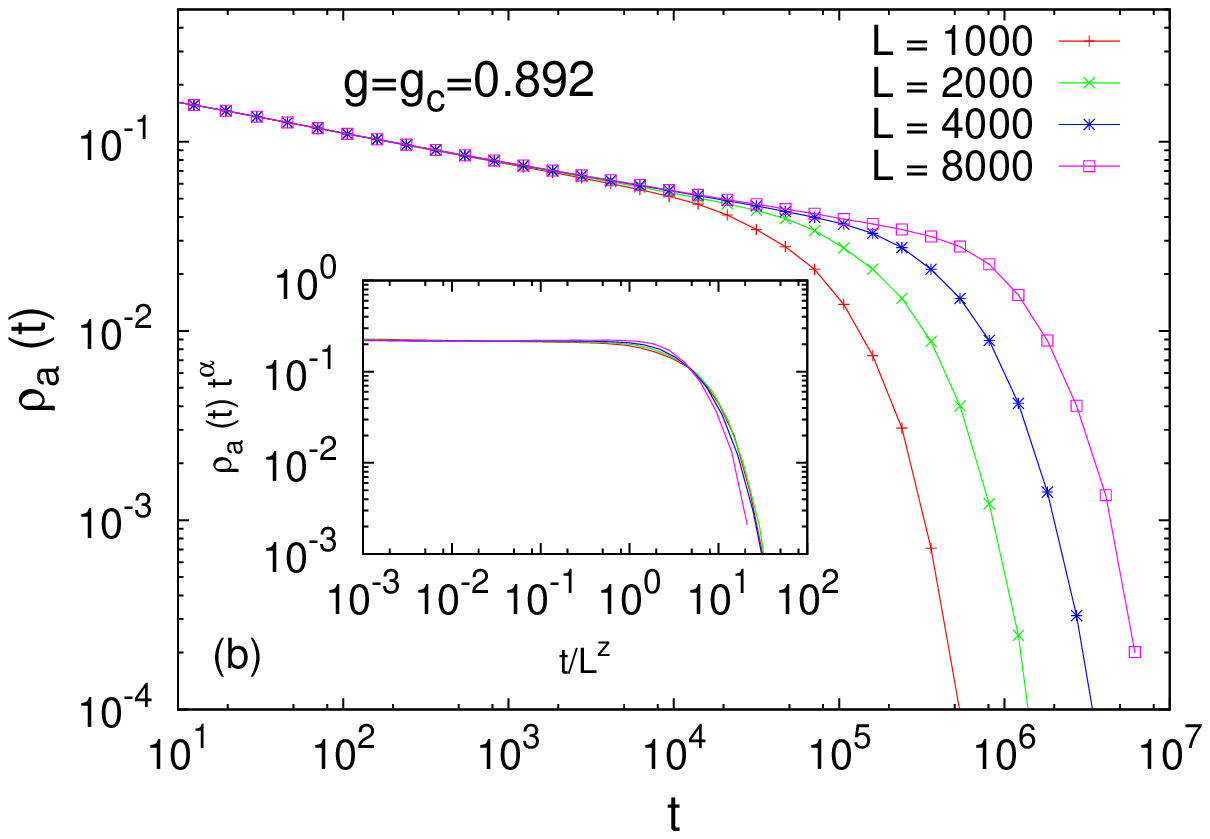}
\includegraphics[width=5.9cm]{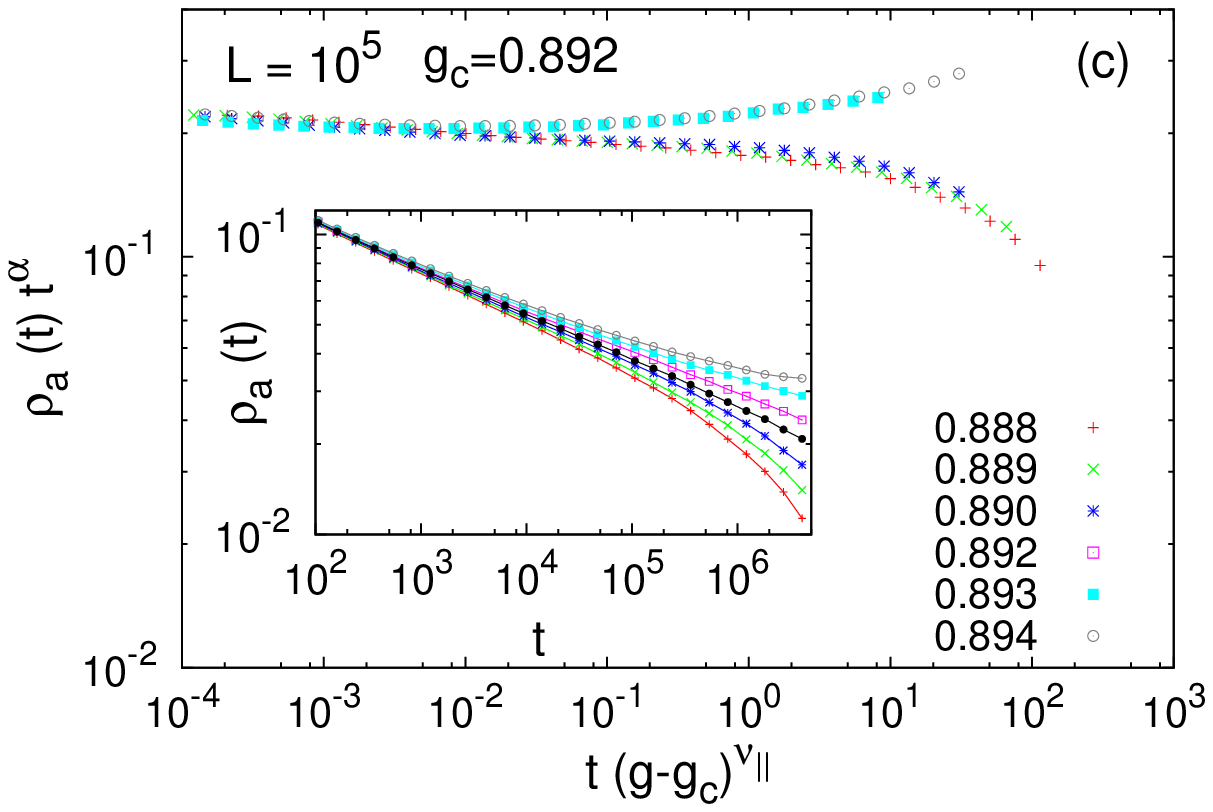}
\caption{Simulation results for the case $p=0$ in $1$-d, $g_c = 0.892 \pm 0.001$.
(a) Variation of steady state density $\rho_a$ of active sites versus $g-g_c$, fitting to $\beta=0.42 \pm 0.01$.
The inset shows the variation of $\rho_a$ with density $g$.
(b) relaxation to absorbing state near critical point for different system sizes $L$,
the inset showing the scaling collapse giving estimates of critical
exponents $\alpha=0.15 \pm 0.01$ and $z=1.40\pm 0.02$. 
(c) Scaling collapse of $\rho_a(t)$. 
The inset shows the variation of $\rho_a(t)$ versus time $t$ for different densities $g$.
The estimated critical exponent is $\nu_\parallel = 1.90 \pm 0.02$.
The simulations are done for linear chains of size $L$ ($=N$).
}
\label{fig:1d}
\end{figure*}



\subsection{Model B}

\subsubsection{Mean field case}
\noindent For the mean field case, we have studied for $N=10^6$, averaging over 
$10^3$ initial condition. We numerically investigate the phase diagram  and the universality 
classes of the transition. In mean field case, the phase boundary seems to be linear 
starting  $g_c=1/2$ for $p=0$ and ending at $g_c=1$ for $p=1$~(Fig.~\ref{fig:pg}),
obeying $g_c =\frac{1}{2}(1+p)$.
In this case,  for $p=0$, we find the critical point 
to be  $g_c=1/2$, and this is similar to the fixed energy sandpiles~\cite{Vespignani}. 
The critical exponents are the same along the phase boundary and they match with those of model A.


\subsubsection{Lattice cases}
\noindent We studied the same dynamics in 1-d and 2-d. For a linear chain in 1-d, 
here also  we took $N=L=10^4$ and average over $10^3$ initial condition. 
For 2-d we took $1000\times1000$ square lattice with $L=1000$ and averaging 
over $10^3$ initial conditions.
\begin{enumerate}[(a)]
 \item 
For $1$-d, for the case $p=0$, we find $g_c = 0.89 \pm 0.01$, with
$\beta = 0.42 \pm 0.01$, $z =1.55 \pm 0.02$, $\nu_\parallel=1.90 \pm 0.02$ and 
$\alpha=0.16 \pm 0.01$ (Fig.~\ref{fig:1d}).
The phase boundary in ($g,p$) is nonlinear: it starts from 
$g_c = 0.89 \pm 0.01 $ at $p=0$~(Fig.~\ref{fig:1d}) to $p = 0.43 \pm 0.03$ at $g=1$~(Fig.~\ref{fig:pg}). 
Thus, one can independently define a model at unit density ($g=1$) and calculate the critical 
probability $p_c$ for which the system goes from an active to an absorbing phase. 

\item 
For $2$-d, for the case $p=0$, we find $g_c = 0.683 \pm 0.002$, with
$\beta = 0.67 \pm 0.02$, $z =1.55 \pm 0.02$, $\nu_\parallel=1.20 \pm 0.03$ and 
$\alpha=0.42 \pm 0.01$.
The phase boundary seems nonlinear, 
from $g_c = 0.683 \pm 0.002$ for $p=0$~(Fig.~\ref{fig:pg}) extending to $g_c=1$ at $p=1$.
\end{enumerate}


\begin{table}[h]

\begin{tabular}{|l|l|l|c|c|c|}
\hline
 & & Model A & Model B & Manna   \\ \hline
\multirow{3}{*}{$\beta$} & 1D  &  & 0.42 $\pm$ 0.01 & 0.382 $\pm$  0.019  \\
 & 2D & 0.68 $\pm$ 0.01  & 0.67 $\pm$ 0.02  & 0.639 $\pm$ 0.009 \\
 & MF & 0.98 $\pm$ 0.02 & 0.99 $\pm$ 0.01  & 1\\
 \hline
\multirow{3}{*}{$z$} & 1D  &  & 1.55 $\pm$ 0.02  & 1.393 $\pm$ 0.037\\
 & 2D  &  1.65 $\pm$ 0.02 & 1.55 $\pm$ 0.02 & 1.533 $\pm$  0.024 \\
 & MF & 2.00 $\pm$ 0.04 &  2.0 $\pm$ 0.04  & 2\\
\hline
\multirow{3}{*}{$\alpha$} & 1D  &  & 0.16 $\pm$ 0.01 & 0.141 $\pm$ 0.024 \\
 & 2D & 0.42 $\pm$ 0.01 & 0.42 $\pm$ 0.01 & 0.419$\pm$ 0.015 \\
 & MF & 1.00 $\pm$ 0.01 & 1.00 $\pm$ 0.01  & 1\\
\hline
\multirow{3}{*}{$\nu_{\parallel}$} & 1D  & & 1.90 $\pm$ 0.02 & 1.876 $\pm$ 0.135 \\
 & 2D & 1.24 $\pm$ 0.01 & 1.20 $\pm$ 0.03 & 1.225 $\pm$ 0.029 \\
 & MF & 1.00 $\pm$ 0.01 & 1.00 $\pm$ 0.01 & 1\\

\hline
\end{tabular}
\caption{Comparison of critical exponents of this model with those of the conserved Manna model \cite{Lubeck:2004}.}\label{table:exponents}
\end{table}

\section{Analytical treatment of the models in mean field case}
\label{sec:ana}
\begin{figure}
\includegraphics[width=8.cm]{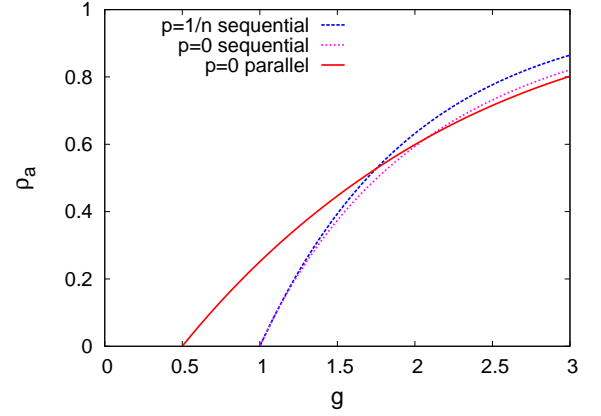}
\caption{Behavior of order parameter for sequential and parallel dynamics. For sequential dynamics, $g_c=1$
for both models. For the model with $p=0$, the parallel dynamics gives $g_c=\frac{1}{2}$.}
\label{fig:ana}
\end{figure}

\noindent In the mean field case the particles can jump from a site 
to {\em any} other, all with equal probability, or, 
equivalently, the underlying graph is fully connected.
From the theory of zero range processes we know that the stationary probability distribution
is factorized in terms of single site quantities.
If $\rho_n^i$ is the probability that a site $i$ has $n$ particles, 
the average rate of outgoing particles is $\langle u^i \rangle = \sum_n \rho_n^i v(n) n$.
By symmetry, we can work under the hypothesis that $\rho_n^i = \rho_n$ $\forall i$, we
have simply $\langle u^i \rangle = \langle u \rangle = \sum_n n v(n) \rho_n$, 
that at stationarity is equal to the average number of incoming particles, since the density is fixed. 
A sequential dynamics would consist of sequences of one-particle jump events and
the master equation has the form:
\begin{eqnarray}
&& \dot{\rho}_k(t)  = 
-(\langle u \rangle + k v(k) )\rho_k(t) \nonumber \\
&& + \theta(k) \langle u \rangle \rho_{k-1}(t) 
+ (k+1) v(k+1) \rho_{k+1}(t). \nonumber \\
\end{eqnarray}
Multiplying by $s^k$, summing over $k$ we have 
the equation for the characteristic function $G(s)=\sum_k \rho_k s^k$:
\begin{equation}
\dot{G} = -(1-s) ( \langle u \rangle G(s) - \sum_{n=1}^\infty n v(n) \rho_n s^{n-1}) 
\end{equation}
from which we have a self-consistent formula for the stationary solution:
\begin{equation}
\label{seq}
G(s) = \frac{1}{\langle u \rangle} \sum_{n=1}^\infty n v(n) \rho_n s^{n-1}.
\end{equation}
For the case (A) the solution of Eq.(\ref{seq}) is
\begin{equation}
G(s) = \theta(1-g)(1-g+sg) + s\theta(g-1)e^{-(g-1)(1-s)} 
\end{equation}      
For the case (B) the solution of Eq.(\ref{seq}) is
\begin{equation}
\label{eq:gs}
G(s) = \theta(1-g)(1-g+sg) + \theta(g-1)  \frac{e^{sx}-1}{x}(g-x), 
\end{equation} 
where $x$ is the solution of $g = \frac{xe^x}{e^x-1}$.
In both cases below $g_c=1$ the activity is zero, i.e. the system falls into the absorbing state till this is present.
The order parameter or the fraction of active sites as a function of $g$ is depicted in Fig.~\ref{fig:ana}. 
On the other hand, the parallel dynamics of the model B with $p=0$ is particularly simple to analyze, 
since the number of particles on top of each site follows the discrete evolution equation:
\begin{equation*}
n_{t+1} = \left\{
\begin{array}{rl}
n^i & \text{if } n(t) \neq 1,\\
n^i+1 & \text{if } n(t) = 1.
\end{array} \right.
\end{equation*}
Then, we have the self consistent equation for the stationary state:
\begin{equation}
\rho_n = e^{-g+\rho_1}\left[ (1-\rho_1) \frac{(g-\rho_1)^n}{n!} + \theta(n)\rho_1  \frac{(g-\rho_1)^{n-1}}{(n-1)!}\right].
\end{equation}
From we which we can work out the characteristic function 
\begin{equation}
G(s) = (1-\rho_1 + s \rho_1) e^{-(g-\rho_1)(1-s)}
\end{equation}
From $G'(0)=\rho_1$, we end up with an equation for $x = g-\rho_1$, that, apart from the solution $x=0$, has the form
\begin{equation}
\label{eq:14}
g = \frac{x(1+xe^{-x})}{1-e^{-x}+xe^{-x}}.
\end{equation}
Finally, the order parameter can be calculated consistently
once we know $G(s)$ from $x$ (Eq.~(\ref{eq:gs})), since $\rho_a = 1-G(0)-G'(0)$ and
its behavior as function of $g$ is depicted in Fig.~\ref{fig:ana}.

\subsection{Approximate analysis of the critical point and faster-is-slower effect}
\noindent To get insights on the value of the critical point as well as on the time to reach 
a frozen configuration below $g=1$, we will analyze the dynamical stability 
of the frozen phase probing it with a simple perturbation of the form:
\begin{eqnarray}
\label{perturb}
\rho_0 = 1-g + \delta \\ 
\rho_1 = g-2\delta \\
\rho_2 = \delta 
\end{eqnarray}
i.e. we pick a fraction $2\delta$ of particles and move them to already filled sites, 
neglecting the case in which some of them choose the same site.
This perturbation evolves according to the equation~\cite{note}
\begin{eqnarray}
\delta(t+1) =\delta(t) - (1-g + \delta(t))(1-e^{-2v(2)\delta(t)}) \nonumber \\
 + \delta(t)v(2)^2 e^{-2v(2)\delta(t)}  
\end{eqnarray}
whose solution scales, for large enough times, like
\begin{equation} 
\delta(t) \propto \left[2v(2)\left(1-\frac{v(2)}{2}-g\right)\right]^t
\end{equation}
with a relaxation time
\begin{equation}
\tau = -\frac{1}{\log\left[1-2v(2)(1-\frac{v(2)}{2}-g)\right]} 
\end{equation}
that diverges in $g_c = 1 - \frac{v(2)}{2}$.
This value coincides in a very good approximation with the numerical values  of the critical point 
both for for the model A, $g_c=3/4$ and B, $g_c = \frac{1}{2} (1+p)$.

\begin{figure}
\includegraphics[width=7.0cm]{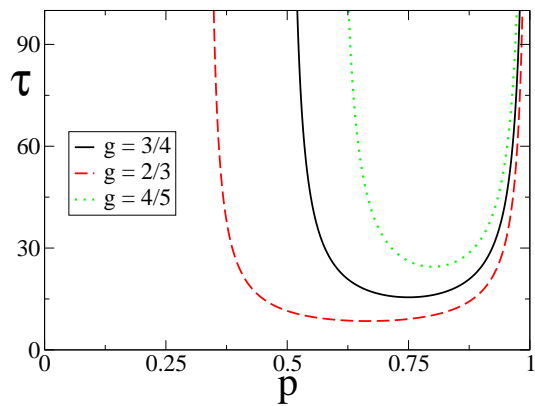}
\caption{Relaxation time $\tau$ as a function of the waiting probability $p$ for the model B for a few values of $g$.
It diverges at $p=1$ and $p=2g-1$ and has an optimal minimum at $p^*=g$,
when $\tau= \tau^*$.}
\label{fastslow}
\end{figure}

Moreover, for the model B the relaxation time at fixed density as a function of $p$ 
\begin{equation} 
\tau(p) = -\frac{1}{\log\left[1-2(1-p)\left(\frac{1+p}{2}-g\right)\right]} 
\end{equation}
has a minimum, optimal value $\tau^* = -\frac{1}{\log[g(2-g)]}$ in $p^*=g$,
as it is shown in Fig.~\ref{fastslow}.
This is a very simple example of the faster-is-slower effect, by which agents who choose a strategy which imply longer waiting times 
allow for a faster collective coordination.
Note that waiting can be rational at the individual level,  
if an agent expects that others are leaving her site. 


\subsection{Analysis of the finite size effects on the time to reach the absorbing state.}
\begin{figure}
\includegraphics[width=7.2cm]{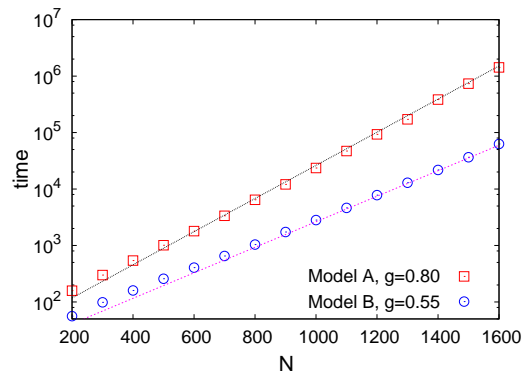}
\caption{Exponential dependency of timescale to reach the absorbing state, with the system size in mean field cases for Model A at $g=0.55$ and Model B at $g=0.80$. The data are averaged over $10^4$ realizations.}
\label{frozen}
\end{figure}

\noindent Let's call  $\pi$ the weight of the frozen configuration for given  $N$ sites and $K=g N$ particles, 
its inverse can be an estimator of the time required to reach the absorbing state.
We have:
\begin{equation}
\pi=\sum_{frozen} P(\overrightarrow n) 
\end{equation}
Since the process is zero-range, at stationarity $P(\overrightarrow n)=\prod_i^N p_i(n_i)$, then, if the graph is homogeneous we can leave the dependency on the site to get 
\begin{equation}
 \pi= \binom{N}{K} \rho_1^K \rho_0^{N-K}
\end{equation}
 Using  Stirling approximation we have finally
\begin{equation}
 \log \pi \simeq -N \left[g\log\frac{g}{\rho_1}+(1-g) \log \left(\frac{1-g}{\rho_0}\right) \right]
\end{equation}
i.e. it is exponentially decreasing with the system size. 
The Fig.~\ref{frozen} shows some typical numerical results for the average time to reach an absorbing configuration as a function of the system size. 

\section{Summary and discussions}
\label{sec:sum}

\noindent The dynamics of social systems is recently gaining more and more insights 
borrowing concepts and techniques from statistical mechanics and by developing some toy models. The toy model paradigm~\cite{Marsili:2007} helps us to understand complex emergent behavior of such socio-economic systems.
In this paper we show how a crowd dynamics in a  resources allocation game 
gives rise to a phase transition between an active 
and a frozen phase, as the density varies. In this respect, we have defined and studied a class of models, 
where $gN$ agents compete among themselves to get the best service 
from $N$ restaurants of same rank, generalizing the `Kolkata Paise Restaurant' problem.
In the original problem, where density $g = 1$
was far from its critical value $g_c$, the relaxation time
$\tau$, given by Eqn.~(\ref{scaling-relation}), never showed any $L = N^{1/d}$ dependence. 
We recast these models in terms of zero-range interacting particles 
in order to get analytical insights on the systems' behavior.  
As long as $g\le 1$, absorbing frozen configurations are present, and 
that can be reachable or not, depends on the underlying dynamics.
We found the existence of a critical point $g_c$ above which the agents  
are unable to find frozen configurations.
In the case in which the agents are moving if and only if they are unsatisfied (model B) with $p=0$,   
they fail to reach satisfactory configurations if the density is above $g_c=1/2$.
Strategies where agents wait longer (higher $p$) speed up the convergence, 
increasing $g_c$ and decreasing the time to reach saturation configurations
(faster-is-slower effect). 
We investigated numerically the phase transition in finite dimensions finding a good agreement
with the exponents of stochastic fixed-energy 
sandpile~(Table.~\ref{table:exponents})~\cite{Vespignani,Manna:1991,Lubeck:2004}.
Thus, we have a simple model for resource allocation, which is solvable, and shows a variety of
interesting features including phase transitions as in well known models.
Further investigations are needed in order to understand the emergence of waiting strategies, 
i.e. how they could depend on agents' beliefs or learning processes, as well as it would 
interesting the study the case of heterogeneous agents (with different strategies)
and/or restaurants (with different ranks).





\begin{thebibliography}{99}
\bibitem{note}
We neglect terms coming from sites with more than two particles. 
The stationary solutions worked out previously reveal that even
slightly above the critical point we  have $\rho_2 \propto (g-g_c)$, whereas in general, for $k>2$, $\rho_k \propto (g-g_c)^{k-1}$.
\end{thebibliography}
\end{document}